\documentclass[prl,twocolumn,superscriptaddress,groupedaddress]{revtex4}  

\usepackage{graphicx}  
\usepackage{xcolor}  
\usepackage{fancyhdr}
\usepackage{dcolumn}   
\usepackage{bm}        
\usepackage{amsmath,amssymb,siunitx}
\usepackage[version=3]{mhchem}
\usepackage{upgreek}
\usepackage{lastpage}
\usepackage{xr}
\newcommand*\chem[1]{\ensuremath{\mathrm{#1}}}

\begin{document}

\title{Motile bacteria in a critical fluid mixture}

\author{Nick Koumakis}
\author{Cl\'emence Devailly}
\author{Wilson C. K. Poon}

\affiliation{SUPA and School of Physics \& Astronomy, The University of Edinburgh, James Clerk Maxwell Building, Peter Guthrie Tait Road, Edinburgh EH9 3FD, Scotland, UK}


\begin{abstract}
We studied the swimming of \textit{Escherichia coli} bacteria in the vicinity of the critical point in a solution of the non-ionic surfactant \chem{C_{12}E_{5}} in buffer solution. In phase contrast microscopy, each swimming cell produces a transient trail behind itself lasting several seconds. Comparing quantitative image analysis with simulations show that these trails are due to local phase re-organisation triggered by differential adsorption. This contrasts with similar trails seen in bacteria swimming in liquid crystals, which are due to shear effects. We show how our trails are controlled, and use them to probe the structure and dynamics of critical fluctuations in the fluid medium.
\end{abstract}

\maketitle


\section{Introduction}

{\color{black} Active matter has attracted the attention of diverse investigators in recent years \cite{Ramaswamy2010}. Self-propelled micro-swimmers, natural or synthetic, are instances of active matter \cite{FermiColloids}. Their lack of time-reversal symmetry promises new physics \cite{CatesReview2012}, and novel kinds of active self assembly can be envisaged \cite{Cates2016,Arlt2018}. A significant theme in active matter research concerns micro-swimmers in complex passive environments, both static and dynamic. Thus, restricting our attention to natural micro-swimmers, we find recent studies on motile bacteria in concentrated polymer solutions  \cite{martinez2014flagellated}, colloidal structures \cite{brown2016swimming}, emulsion drops \cite{Vladescu2014,Lushi2014}, liquid crystals \cite{Zhou2014,Abbott2014}, gels \cite{Croze}, arrays of static, asymmetric obstacles \cite{Galajda}, and intricate microfluidic devices \cite{Wioland2016}. 

One complex passive medium that is yet to be explored in this context is a critical fluid. Critical binary fluids have been used to generate self propulsion in synthetic particles via the manipulation of critical fluctuations by surface asymmetries \cite{buttinoni2012active,Buttinoni2013}. However, the way these fluctuations may interact with micro-swimmers has not been probed to date. 

Near the critical point, the correlation length diverges according to $\xi \propto \epsilon^{-\nu}$, with $\epsilon = |(T-T_c)|/T_c$ measuring how close the temperature $T$ is to the critical temperature $T_c$. For small enough $\epsilon$, $\nu \approx 0.630$ for binary fluids \cite{Roy2016}; for larger $\epsilon$, $\nu = \tfrac{1}{2}$. The characteristic decay time of fluctuations also diverges: $\tau \propto \epsilon^{-\nu z}$, where $z = d + x_\eta$ in $d$ dimensions, and $x_\eta \approx 0.068$ \cite{Roy2016}. We expect these fluctuations to decay with a diffusivity that scales as $D \sim \xi^2/\tau \sim \epsilon^\phi$ with $\phi = \nu(d-2 + x_\eta) > 0$ \cite{Roy2016}, so that $D \rightarrow 0$ as $\epsilon \rightarrow 0$. 

For an active particle of characteristic dimensions $L$ swimming in a critical binary mixture at speed $v$, the characteristic advective and diffusive time scales are $t_{\rm a} \sim L/v$ and $t_{\rm d} \sim L^2/D$, so that the P\'eclet number Pe~$= t_{\rm d}/t_{\rm a} \sim Lv/D$ diverges near criticality. We may therefore expect non-trivial effects in such a system. 

We explore this possibility using motile {\it Escherichia coli}, a model active colloid \cite{schwarz2016escherichia} with body length $L \gtrsim \SI{2}{\micro\meter}$ typical speed $v \sim \SI{20}{\micro\meter\per\second}$, in a critical mixture of the non-ionic surfactant \ce{C12E5} and water. In this mixture, $\xi \sim \SI{0.3}{\micro\meter}$ and $\tau \gtrsim \SI{0.01}{\second}$  at a readily accessible $\epsilon \sim 10^{-4}$ \cite{hamano1985dynamics,hamano1995static,sinn1992}, giving $D \lesssim \SI{10}{\square\micro\meter\per\second}$ and Pe~$\sim Lv\tau/\xi^2 \gtrsim 1$. Under these conditions, we find that the motile bacteria `paint' a transient pattern of trails in the critical fluid reminiscent of the process whereby Jackson Pollock created his iconic canvasses \cite{Pollock}.

 

Visually similar trails seen in bacteria swimming in liquid crystals \cite{Zhou2014} are due to shear effects. Our trails, on the other hand, relate to the way the swimmers modify the structure and dynamics of the critical fluctuations. This effect is not confined to self-propelled particles, but has also been predicted for colloids being dragged through critical fluids \cite{Furukawa2003,Demery2010}, so that our experiments using active colloids provide the first experimental verification of this prediction based on simulations of passive particles. At the same time, our results constitute another example of micro-swimmers acting as a local probe of complex fluids, albeit in a roundabout way \cite{martinez2014flagellated,Zhou2014}.} 



\section{Methodology}

\subsection{Experimental}

We cultured a $\Delta$cheY (non-tumbling) mutant of strain AB1157 in Luria Broth (LB) agar plates. A single colony was then transferred to 10ml of liquid LB to be incubated at $30^{\circ}$C overnight. Then 350 $\mu$L of this culture was transferred into 35mL of Terrific Broth (TB) to grow for 4h, reaching the mid-exponential phase. Cells were then washed 3 times in a motility buffer (MB: 6.2 mM \ce{K2HPO4} (Sigma-Aldrich), 3.8 mM \ce{KH2PO4} (Fisher Chemical) and 0.1mM EDTA (Sigma Aldrich)) using a filter unit and 0.45$\mu$m filter paper. We then transfered the bacteria into our critical fluid. Our final critical fluid consisted of a mixture of $1.6$ wt.\% pentaethylene glycol monododecyl ether, \chem{C_{12}E_{5}}, (Sigma Aldrich), a non-ionic surfactant, the motility buffer (MB) and $1.5$ mM glucose. The glucose promotes stable swimming for several hours \cite{schwarz2016escherichia}, while the concentration of \ce{C12E5} was non-harmful to {\it E. coli} swimming. The final bacterial suspension, having an optical density (OD) between $0.05$ and $0.1$, was loaded into $50\mu$m $\times$ $0.5$mm rectangular capillaries (CM Scientific). 

We used a lower critical solution temperature (LCST) critical mixture. This critical mixture was chosen because of its compatibility with bacterial swimmers, its large microscopic correlation length and flat phase boundary near criticality, the latter giving a large window for observing critical fluctuations \cite{hamano1995static}. The solution was stored away from light. The precise phase boundary varied between experiments due to sample degradation by dissolved \ce{CO2} and photodissociation (minimised by storage in the dark) as well as external temperature variations. Therefore, $T_c$ was determined in each experiment separately, by warming a sample up from the single-phase to the two-phase region through the critical point. The trail measurements were then taken immediately using a sample in a fresh capillary to avoid any hysteresis effects from improper remixing \cite{Martinez2017Clem}.

Using phase contrast microscopy (Nikon $20\times$ ph1 and $60 \times$ ELWD ph3 objectives) near a glass surface and a 16-bit high-sensitivity camera (Orca Flash, Hamamatsu), we imaged bacteria and their trails. An INSTEC temperature stage (mK1000) kept the sample temperature stable to $\pm \SI{0.01}{\celsius}$. The samples were taken to $\SI{0.1}{\celsius}$ below $T_c$, while $T_c$ was determined for each experiment with an error of $\pm \SI{0.05}{\celsius}$. Measurements at different $\epsilon$ were then carried out by successively decreasing the temperature. Ten $\SI{40}{\second}$ movies at 100 frames per second were taken at each temperature after stabilising for \SI{5}{\minute}.

We also extracted the diffusion coefficient $D^{\rm Bulk}$ of the density fluctuations using differential dynamic microscopy (DDM) following a published protocol \cite{Giavazzi2016}. Movies of the binary mixture without bacteria at different temperatures were taken ($60 \times$ ELWD ph3 Nikon objective, $40$s at 100 frames per second, $512 \times 512$ and $4\times $4 binning). This gave access to a wave vector range of $0.05 \leq q \leq \SI{6.6}{\per\micro\meter}$. The extracted differential intermediate scattering functions \cite{Giavazzi2016} exhibited single exponential decay, which we fitted with $A(q)(1-e^{-Dq^2t})+B(q)$, where $t$ is the delay time between two frames. $A(q)$ is determined by the structure factor and form factor of the sample, while $B(q)$ is the measured camera noise. We obtained a value of $D$ for each available $q$ and averaged over $0.5 \leq q \leq \SI{2.4}{\per\micro\meter}$, the region in which reliable correlation functions could be obtained.

\subsection{Simulations}

To further our understanding of the observed phenomenon, we simulated the 2D conserved order parameter (COP) Ising model \cite{newman1999monte}. To model a binary (AB) mixture, the $N \times N$ spins in the COP Ising model are defined to be $s_i = +1$ or -1 if site $i$ is occupied by species A or B. The total energy is $H = \tfrac{J}{4} \sum_{ij} (1 - s_1s_2) = \mbox{constant}\, - J \sum_{ij} s_is_j$ summed over nearest neighbours with $J > 0$. We require conservation of $n = N^{-1} \sum_i \sigma_i$, the normalised (signed) difference in numbers of A and B. The critical point is at $T_c \approx 2.27$ in units of $J/k_B$ and $n_c = 0$. We work at $n_c$, where at $T < T_c$, the system separates into equal amounts of symmetric coexisting A- and B-rich phases, i.e.~$n = \pm \Delta n$. In our Monte Carlo (MC) simulations, $n$-conservation is implemented using  Kawasaki (or nearest-neighbour spin exchange) dynamics \cite{newman1999monte,Kawasaki1966}. 

At each step of our Monte Carlo simulation, we chose a random site and one of its neighbours and calculated the energy change $\Delta E$ for exchanging the two spins. We used the heat-Bath algorithm \cite{newman1999monte}, i.e.~a spin exchange probability of $p=(1+e^{\Delta E/T})^{-1}$. We define $t_s$ as the time when spin exchange attempts = number of cells in the system. Simulations were performed over for of $T = 2.7$ and 2.8 ($T_c \approx 2.27$, $\epsilon=0.19$, $0.23$) \cite{newman1999monte} on $1050\times1050$ and $540\times540$ cell systems. 

Two types of simulations were performed. Initially, in order to mimic a swimming bacterium, we allowed our Ising model simulation to progress at $T=2.8$ ($\epsilon \approx 0.23$), with a Gaussian-shaped energy inclusion (peak value = $5.6 k_BT$, standard deviation = 4 pixels (px)) travelling at a speed of 0.05 px/$t_s$. We extracted the average spin when centred around the energy inclusion, after reaching a steady state. To distinguish the origin of the trail between phase separation and shear mixing a second simulation was performed, where a 'trail' was imposed and let to equilibrate. For local phase separation, we allowed the simulation to equilibrate at $T=2.7$ ($\epsilon \approx 0.19$) and subsequently applied a positive spin trail. In the case of mixing, we constructed initial conditions by imposing a projected 1D spatial variation of a high temperature and simulated until a steady state was reached. The spatial variation for the temperature or spin profile was a step function with a diameter of $64$ sites. A waiting time of $t=100000 t_s$ was found to be sufficient for reaching a steady state in all cases. The new profile was then allowed to equilibrate without external control and the the relaxation dynamics were qualitatively compared to the experiments. We typically averaged results from $> 10$ runs.

\section{Results}

\begin{figure}[t]
\includegraphics[width=0.5\textwidth,clip]{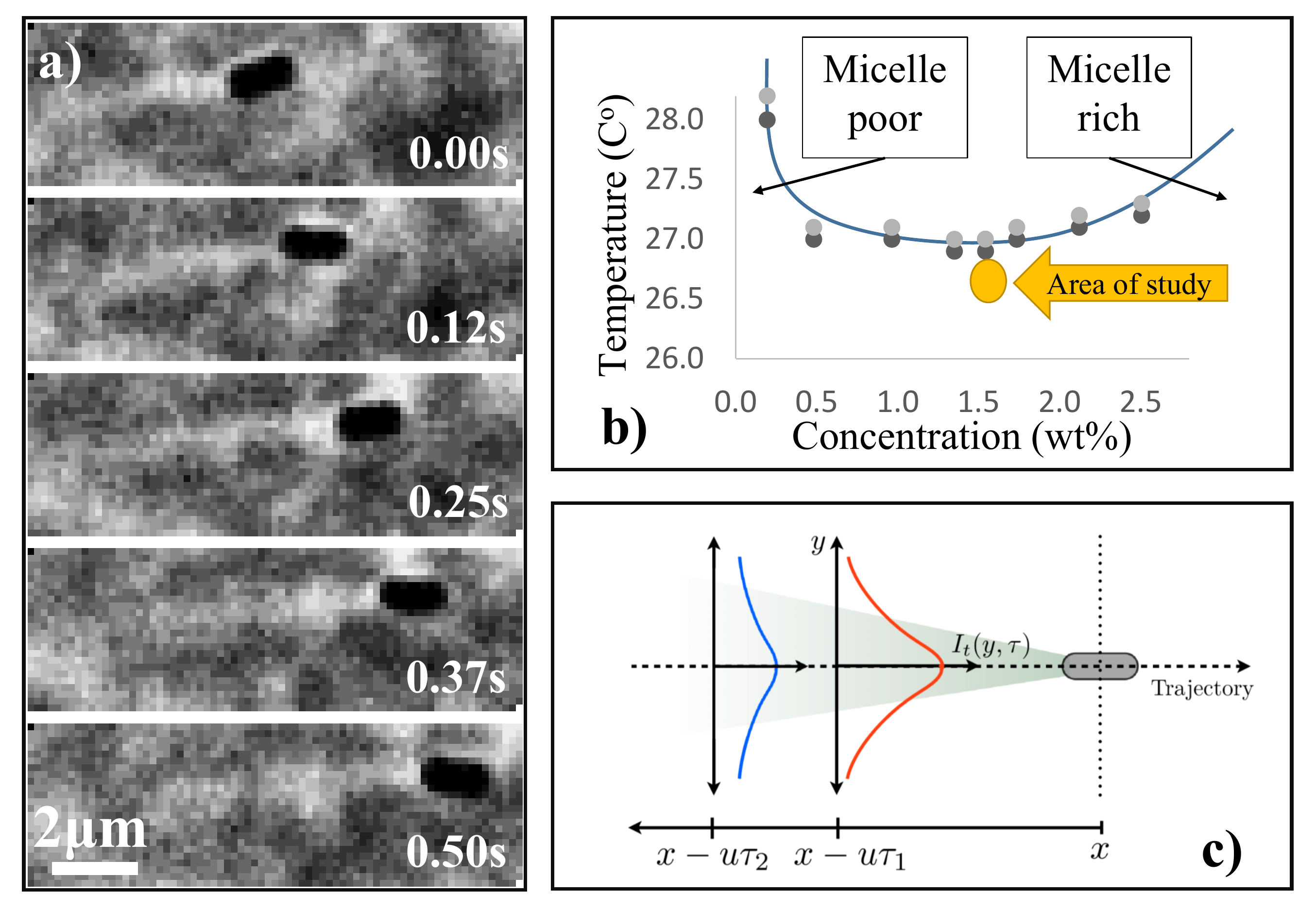}
\caption{\label{fig:Fig1} a) Phase contrast microscopy image snapshots of a single swimming \textit{E. coli} at different times, showing an occurring trail in a phase separating fluid (brighter than the background). The image contrast was increased for clarity. b) Experimental phase diagram of the temperature versus the concentration of \chem{C_{12}E_{5}} in the motility buffer, without bacteria, highlighting the experimental area. Points show the measured upper and lower bounds of the phase separation, while the line is a guide to the eye. c) Schematic of how the trail intensity profiles are extracted from the images.} 
\end{figure}

We observed smooth-swimming mutants of {\it E. coli} dispersed in a solution of \ce{C12E5} in phosphate motility buffer with added glucose, in which cells remain motile for several hours. The solution's phase diagram, Fig.~\ref{fig:Fig1}(b), resembles that of \ce{C12E5} in pure water \cite{hamano1985dynamics,hamano1995static}, but the critical point is shifted from $c = 1.2$wt.\% to $c_0 = 1.6$wt.\% \ce{C12E5} and a lower critical solution temperature (LCST) from $T = \SI{31.8}{\celsius}$ to $T_c = \SI{27.0}{\celsius}$. We monitored cells swimming at $\gtrsim \SI{10}{\micro\meter\per\second}$ in a mixture at $c_0$ and $3 \times 10^{-4} \lesssim \epsilon \lesssim 3 \times 10^{-3}$ (with a variability of $\pm \, \SI{0.05}{\celsius}$ in $T_c$ from sample to sample due to uncertainties in concentration).

Phase contrast microscopy movies revealed a criss-crossed pattern of $\sim \SI{6}{\micro\meter}$ long phase-bright trails behind swimming cells, Fig.~\ref{fig:Fig1}(a). In our set up, phase bright corresponds to lower than average refractive index. Since \ce{C12E5} has higher index than water \cite{Chen2000}, our trails should be water rich. 

This suggest that {\it E. coli} surfaces have a higher overall affinity for water than surfactant, which gives rise to a water-rich fluid layer next to the cell whose extent scales as $\xi$. Near criticality, it extends mesoscopically into the surrounding fluid. As a cell moves, this layer is advected backwards, forming a phase-bright water-rich trail, which then equalises its concentration with the bulk by diffusion. Consistent with this picture, we observed no trails at the same $\epsilon$ but a lower \ce{C12E5} concentration of $c = 1.1$wt\%~$< c_0$, i.e.~richer in water than criticality, but did observe trails at $c = 2.2$wt.\%.

\begin{figure}[t]
\includegraphics[width=0.45\textwidth,clip]{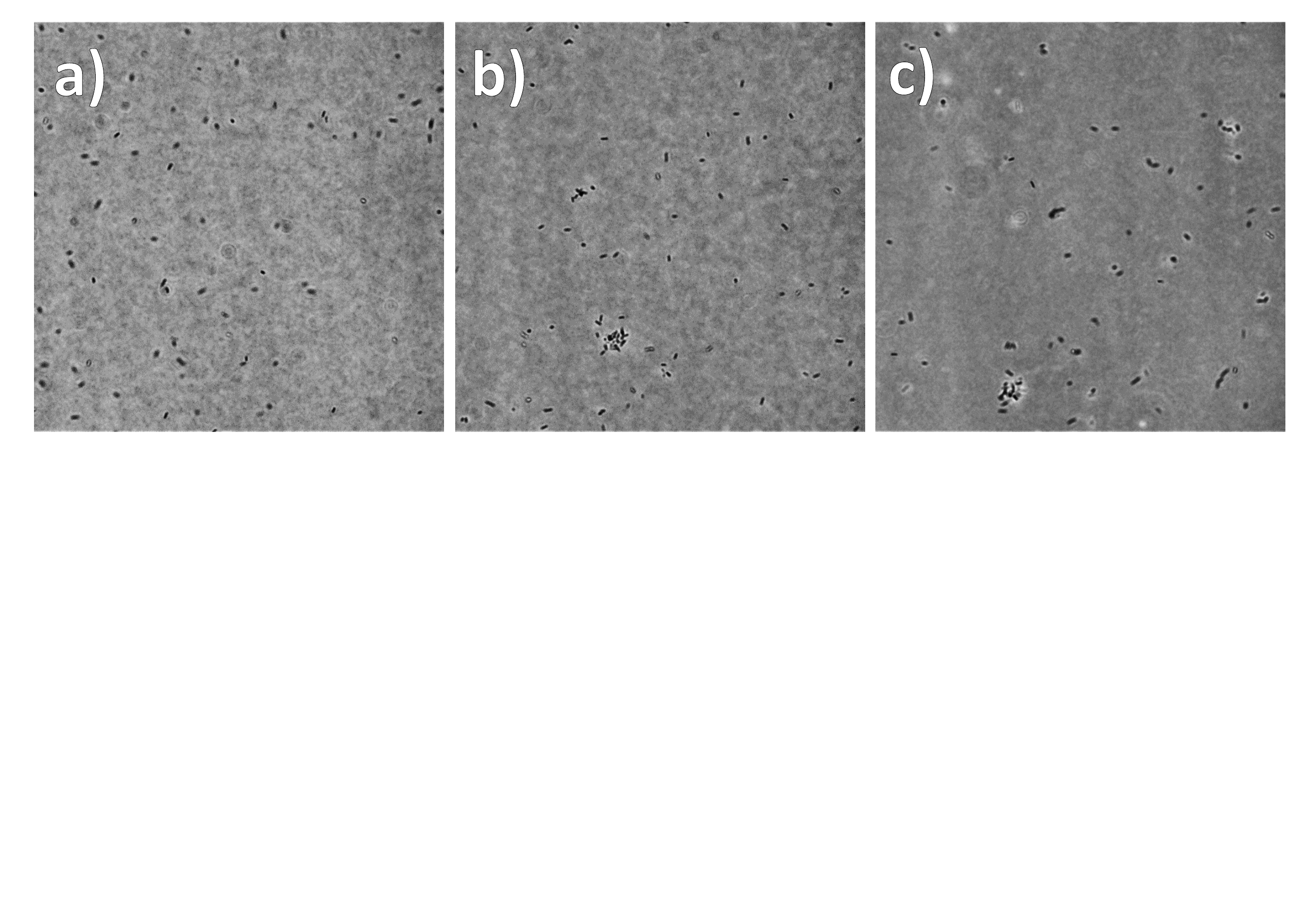}
\vspace{-3.cm}
\caption{\label{fig:Fig2} 
Snapshots of bacteria at a temperature near criticality/phase separation ($\approx-0.1$K), with the concentration of \ce{C12E5} at a) $1.1$wt.\%, b) 1.6\% and c) 2.2\% . Note the appearance of bacterial aggregates for b) and c).
}
\end{figure}

Moreover, we have incidentally observed that in off-critical concentration experiments, bacteria tended to form aggregates whenever trails were visible (with the concentration of \ce{C12E5} at 1.6\% and 2.2\% wt.\%). However, these are anecdotal observations on dilute samples, with aggregates only forming between non-motile bacteria. The resulting clusters appear few and far apart. We show some microscope images of our observations in figure~\ref{fig:Fig2}. This is again consistent with preferential water affinity, which should generate attractive critical Casimir forces \cite{hertlein2008direct,GambassiPhysRevE,NellenEPL,vasilyev2014critical}.

To make a minimal model of trail formation, we consider the bacterium as a source moving at velocity $-\mathbf{u}$ and `emitting' water-rich phase at a constant rate $Q$ per unit time. In the co-moving frame, fluid advects past the cell at the origin at velocity $\mathbf{u}$, and the concentration of water-rich phase, $c(\mathbf{r},t)$, satisfies mass conservation:
\begin{equation}
\partial_t c + \nabla \cdot \mathbf{j} = Q \delta(\mathbf{r}).
\end{equation} 
Substituting the diffusive and advective fluxes $\mathbf{j}_D = -D \nabla c$ and $\mathbf{j}_A = c\mathbf{u}$ (so that $\mathbf{j} = \mathbf{j}_D + \mathbf{j}_A$) and assuming incompressibility ($\nabla \cdot \mathbf{u}=0$) gives 
\begin{equation}
\partial_t c - D\nabla^2 c + \mathbf{u}\cdot \nabla c = Q \delta(\mathbf{r}), \label{general}
\end{equation}
an advection-diffusion equation, which has been used to model pheromone spreading from a moving insect \cite{ecology}. The neglect of hydrodynamics is plausible {\it a priori} because of the dipolar or higher order flow field around a bacterium \cite{DrescherPuller}, and justified {\it a posterori} by fit with experiment. 

In the steady state, $\partial_t c = 0$, and let $\mathbf{u}$ be along $x$. For all times $t$ such that $ut \gg \sqrt{2Dt}$, or $t \gg 2D/u^2$, advection is much faster than diffusion. This requires $x \gg \lambda = 2D/u$. For our $u \gtrsim \SI{6}{\micro\meter\per\second}$ and $0.1 \lesssim D \lesssim \SI{1}{\square\micro\meter\per\second}$ (see later), $\lambda \lesssim \SI{0.3}{\micro\meter}$. 
In the $x \gg \lambda$ (high P\'eclet) limit, we neglect diffusion along $x$, and obtain
\begin{equation}
u\partial_x c = D\left( \partial_{yy} c + \partial_{zz} c\right) + Q \delta(\mathbf{r}).
\end{equation}

For a moving source at $z = h$ above a non-porous wall at $z=0$, the stationary concentration profile is
\begin{multline}
c(\mathbf{r}) = \sqrt{\frac{Q}{2\pi Dx(\tau)}} \exp\left\{-\frac{uy^2}{4Dx(\tau)}\right\} \sqrt{\frac{Q}{2\pi Dx(\tau)}}  \\
 \times \left[ \exp\left\{-\frac{u(z-h)^2}{4Dx(\tau)}\right\} + \exp\left\{-\frac{u(z+h)^2}{4Dx(\tau)} \right\} \right],
\end{multline}
with the second $z$-dependent `image' term enforcing zero flux at $z = 0$. In the lab frame, the distance downstream from the cell in the co-moving frame, $x$, is parametrised by the time interval $\tau$ before the current time $t$ when the cell was at position $x = ut$ downstream from its current position, assuming constant $u$, Fig.~\ref{fig:Fig1}(c), so that the transverse concentration profile at a fixed $z$ is Gaussian:
\begin{equation}
c(y,\tau) = A(\tau) \exp\left\{-\frac{y^2}{4D\tau}\right\}, \label{solution}
\end{equation}

\begin{figure}[t]
\includegraphics[width=0.45\textwidth,clip]{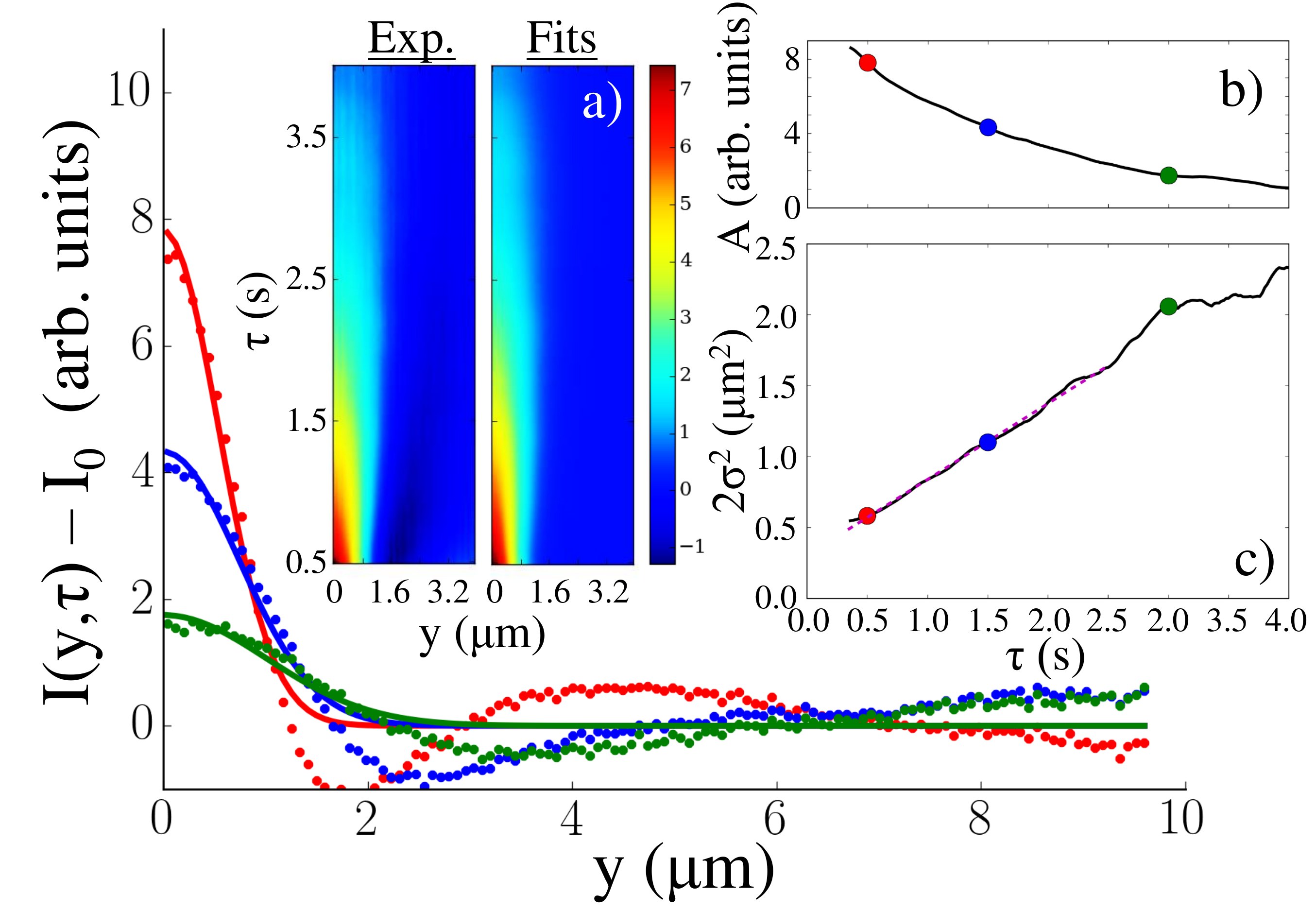}
\caption{\label{fig:Fig3} 
Main: Plot of three intensity profiles at a near-critical temperature ($\epsilon=3.3$ $10^{-4}$, or $T_c - T \approx \SI{0.1}{\celsius}$) after subtraction of the background for $\tau=0.5$ (red), $1.5$ (blue) and $3.0$s (green). Dots are from analysed data, while solid lines are the Gaussian fits to the data. Insets: a) A contour plot of the intensity profiles as a function of time, for the analysed data and for the fitted curves. The height, b), and variance, c), of the Gaussian fits as a function of time, including a linear fit in c), from which a diffusivity can be extracted (Eq.~\ref{eq:Phi}, here $D\approx \SI{0.15}{\square\micro\meter\per\second}$). 
}
\end{figure}

If the intensity in our movies, $I$, is proportional to concentration, then measuring $I_t(y, \tau)$ in individual images, Fig.~\ref{fig:Fig1}(c), can test Eq.~\ref{solution}. In practice, we average over different frames (i.e., over $t$) for each of many cells $i = 1$ to $\sim 10^3$. We plot the result with background subtracted, $\langle I(y, \tau)  \rangle_{t,i} - I_0$, Fig.~\ref{fig:Fig3}, including only data from vigorous swimmers, $v > \SI{6}{\micro\meter\per\second}$ and for $x = v\tau > \SI{6}{\micro\meter}$, the latter  to avoid strong effects from beating flagella. The data can be fitted by Eq.~\ref{solution}:
\begin{align}
\langle I(y,\tau) \rangle_{t,i} = A(\tau) 
\exp \left\{ -\dfrac{y^2}{2\sigma(\tau)^2} \right\} + I_0,
\end{align}
except that the variance has a constant term:
\begin{equation}
2\sigma^2=4 D \tau+2\sigma_0^2, \label{eq:Phi}
\end{equation}
with $D \approx \SI{0.15}{\square\micro\meter\per\second}$ and $\sigma_0 \approx \SI{0.30}{\micro\meter}$, the latter corresponding to a minimum full width of $\lesssim \SI{1}{\micro\meter}$, which is close to the point spread function of our imaging system.

\begin{figure}[t]
\includegraphics[width=0.5\textwidth,clip]{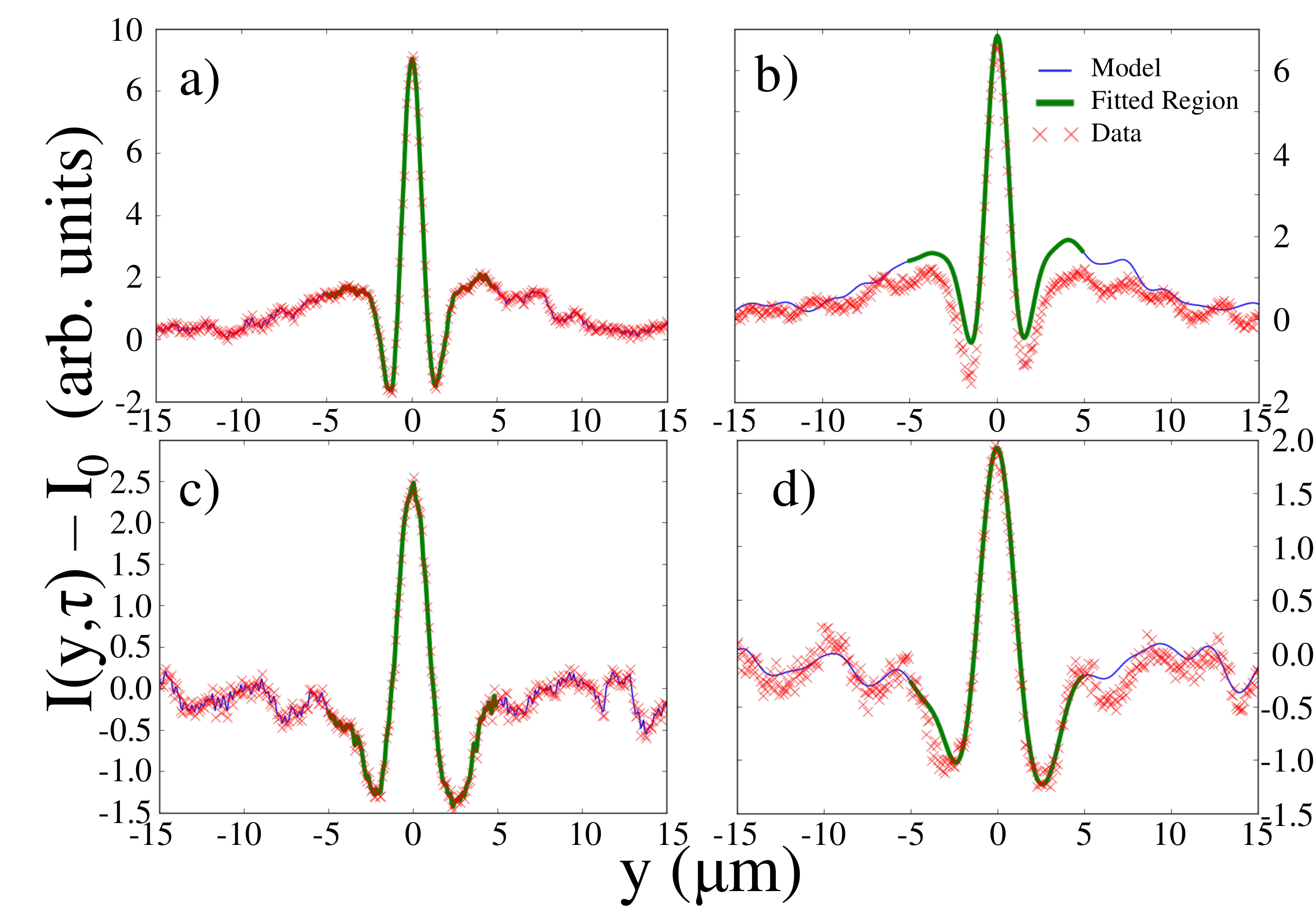}
\caption{\label{fig:Fig4} 
Fitting of the diffusivity for a specific temperature ($\epsilon=6 \times 10^{-4}$), by numerically integrating Fick's second law over the experimental data, showing the experimental data and the fitting. The range of fitting is restricted to $5 \mu m$ from the centre of the track. a) The starting configuration for the model integration, a short time after the trail has been formed ($t_0=0.7$s) and b) the ending configuration ($0.7+0.2$s). c) The starting configuration at a further time ($t_0=1.4$s) after the trail formation and d) the corresponding the ending configuration ($1.4+0.2$s). Note the improvement of the fitting quality between b) and d).
}

\end{figure}  
\begin{figure}[t]
\includegraphics[width=0.5\textwidth,clip]{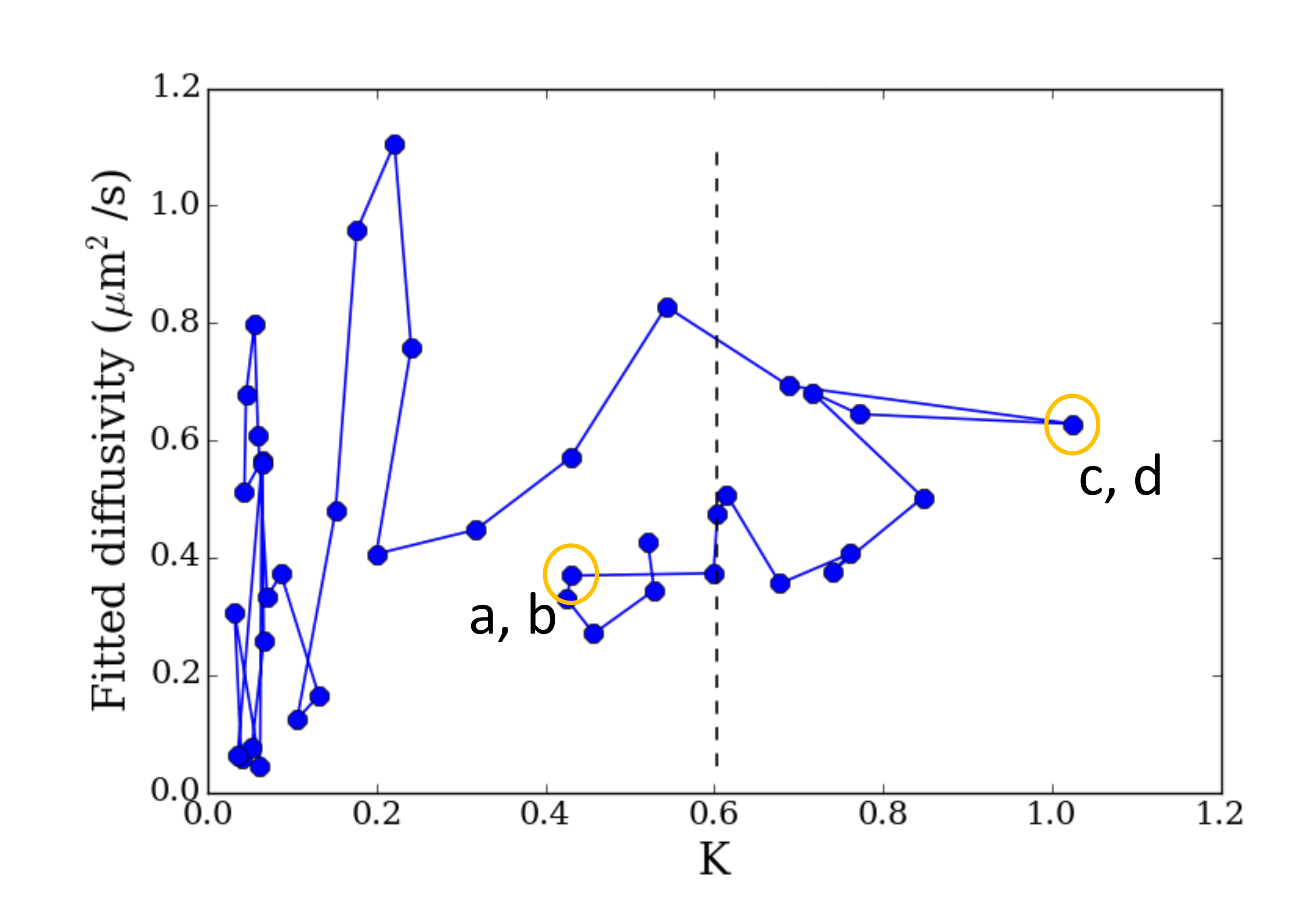}
\caption{\label{fig:Fig5} 
Figure showing the selection of data for the averaging process. The fitted diffusivity (fig. \ref{fig:Fig4}) is plotted against the squared peak value of the trail, multiplied by the reciprocal of the sum of squared deviations between the fit and the experimental data ($K$). The 10 Diffusivities scoring the highest $K$ are then averaged (Noted by the vertical line). The points corresponding to fig. \ref{fig:Fig4} are highlighted.
}
\end{figure}  


This minimal model suggests that the dynamics of the trail is indeed diffusive, Fig.~\ref{fig:Fig3}(c). However, the fitted diffusivity is significantly underestimated, because the trail profile is oscillatory along $y$ and not Gaussian. A more realistic model for the trail profile requires information on initial conditions, which are poorly understood in this experiment. So, instead, we extracted a more realistic diffusivity by numerically calculating how the experimentally-measured initial trail profile at time $t_0$ should spread out diffusively. 

Specifically, we numerically integrate 1D Fickian diffusion, $dI/dt = D d^2 I/dx^2$, for the trail intensity $I(t_0+t)$ and use the experimental profile at time $t_0$ as a starting state. We then compare the time dependent experimental profiles with the simulated ones and minimise their squared differences, using the diffusion coefficient $D$ as a fitting parameter. This method offers a distinctive advantage over using analytical modelling, as it does not require explicit knowledge of the shape of the starting conditions. We fit $D^{\rm Trail}$ over multiple instances of $t_0$, every 0.08s over a duration of 0.2s, examining the fit only 5$\mu$m around the track centre. Our final value of $D^{\rm Trail}$ is averaged over curves which represent meaningful data and are fitted well.

We find that at short times fitting is not ideal (fig. \ref{fig:Fig4}), probably due to phase contrast halo artefacts, as we find that deviations are generally found around the regions of the oscillations. Thus, the data to be averaged is chosen by two criteria: i) the squared height of the peak at the particular $t_0$, denoting the relevance of the data point and ii) the wellness of the fit given by the reciprocal of the sum of the squared difference between the fit and the data. The product of the two produces a non-dimensional parameter $K$ which we can use to rate the fitted diffusivities. A typical example is shown in fig.~\ref{fig:Fig5}. The 10 diffusivities with the highest rating are then used to average the data point of the particular temperature/measurement. 

\begin{figure}[t]
\includegraphics[width=0.45\textwidth,clip]{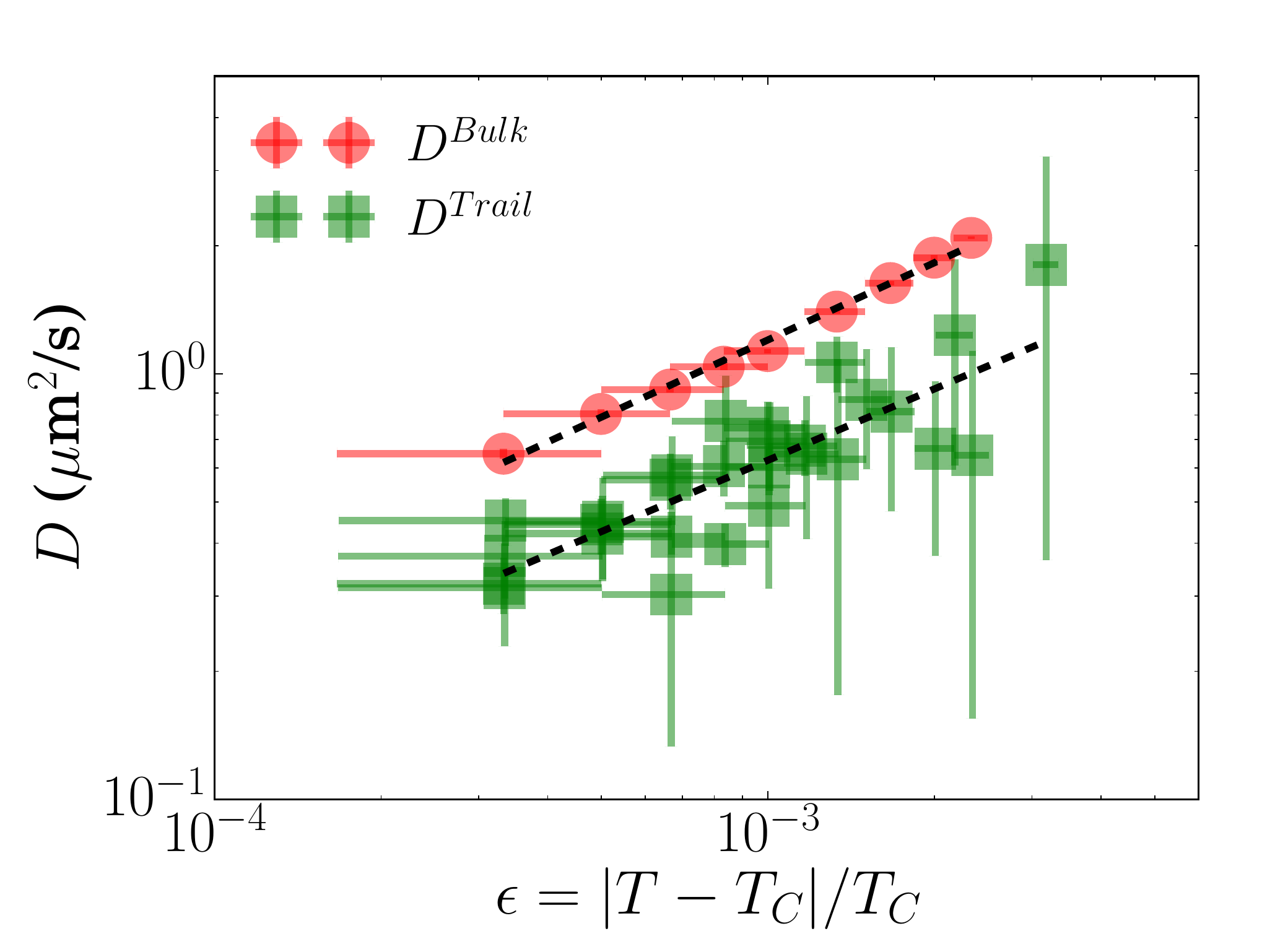}
\caption{\label{fig:Fig6} 
Diffusivity as a function of normalised distance from criticality, $D(\epsilon)$, measured by fitting trail intensity profiles ({\color{green} $\blacksquare$}) and from differential dynamic microscopy ({\color{red} $\bullet$}). Lines are fits to $D \sim \epsilon^\phi$ with $\phi = 0.56 \pm 0.08$ and $0.60 \pm 0.03$ respectively.}
\end{figure}

This procedure was repeated at different temperatures to obtain $D^{\rm Trail}(\epsilon)$, Fig.~\ref{fig:Fig6}. This procedure returns diffusivities that are $\gtrsim 2\times$ those obtained from fitting Gaussians to the trail profiles (data not shown, but one can compare to the value obtained from Fig.~\ref{fig:Fig3}(c)). Theoretically, $D \sim \epsilon^\phi$, where $\phi = {\nu x_D}$ with $x_D = d-2 + x_\eta$ in $d$ dimensions \cite{Roy2016}, so that in 3D, $\phi \approx 0.63 \times 1.068 = 0.67$ (or, using the mean-field $\nu = \tfrac{1}{2}$, $\phi = 0.534$). Fitting $D^{\rm Trail}(\epsilon)$ gives $\phi = 0.56 \pm 0.08$. 

\begin{figure}[t]
\includegraphics[width=0.5\textwidth,clip]{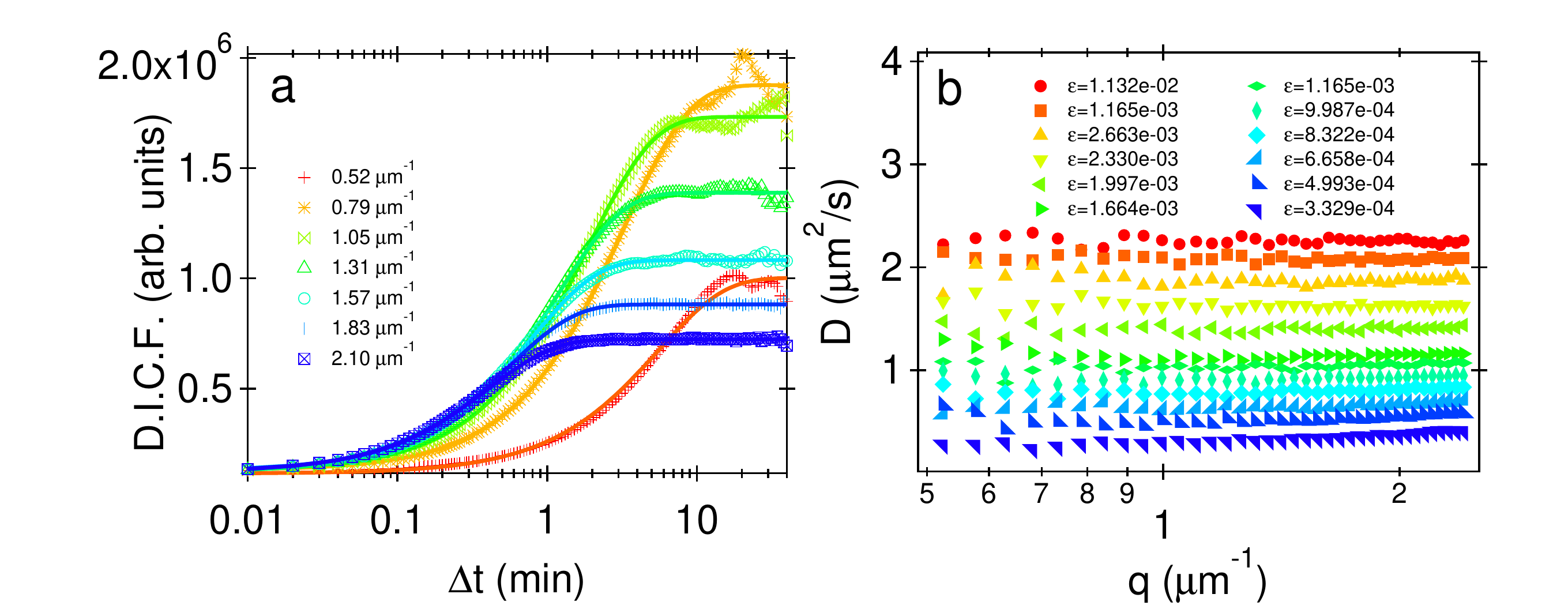}
\caption{\label{fig:Fig7} 
a) Dynamic intermediate correlation function for several wave vectors at a fixed temperature corresponding to $\epsilon=5$ $10^{-4}$. Markers: experimental data; lines: corresponding fit with a single exponential decay. b) Diffusion coefficient extracted from DICF adjustment at each wave vector $q$ for several temperatures. 
}
\end{figure}

Next, we used differential dynamic microscopy (DDM) to measure the diffusivity in the bulk of our mixture~\cite{giavazzi2016equilibrium} (fig.~\ref{fig:Fig7}) as a method of giving another estimate of the diffusivity, $D^{\rm Bulk}(\epsilon)$, Fig.~\ref{fig:Fig6}. The data are less noisy than $D^{\rm Trail}(\epsilon)$, and return a similar $\phi = 0.60 \pm 0.03$. In absolute magnitude, $D^{\rm Bulk} \approx 2 D^{\rm Trail}$. Part of the discrepancy may be due to the difference between the actual 3D profile of the trails, and the 2D phase-contrast projection used for extracting $D^{\rm Trail}$. Despite this discrepancy, the closeness of the $\phi$ exponents from $D^{\rm Trail}$ and $D^{\rm Bulk}$ confirm our hypothesis that the trails dissipate diffusively.

\begin{figure}[t]
\includegraphics[width=0.5\textwidth,clip]{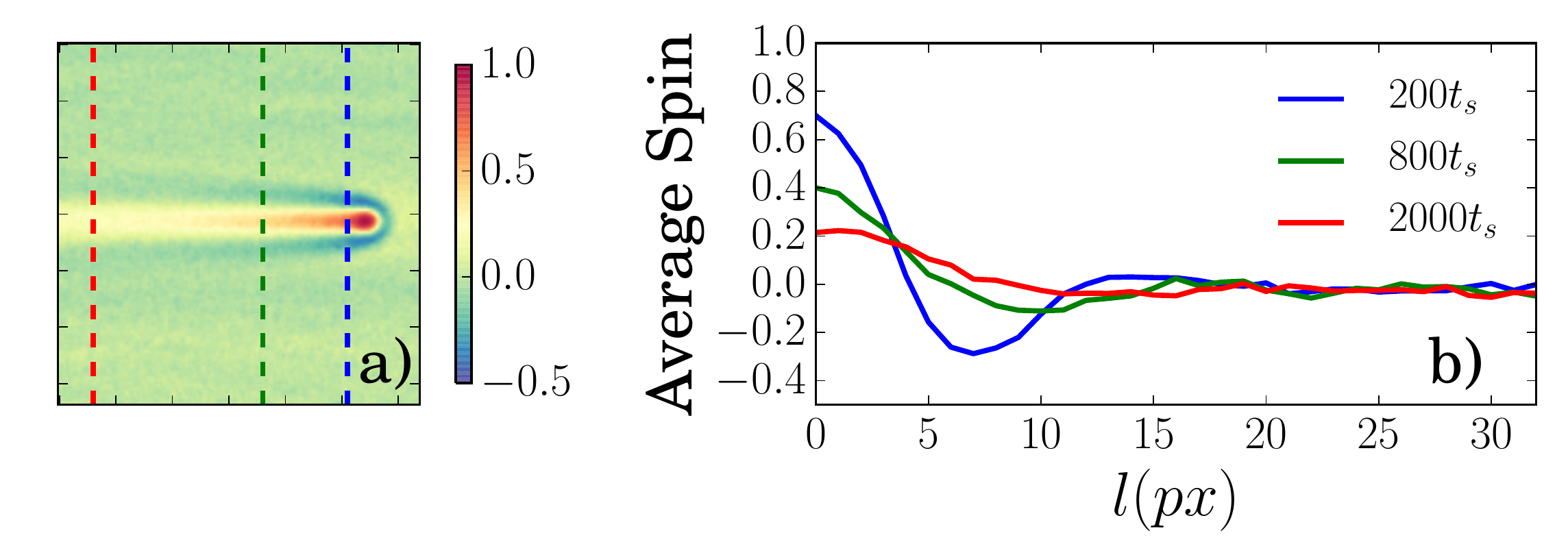}
\caption{\label{fig:Fig8} 
a) Ising model simulation at $T=2.8$ ($\epsilon \approx 0.23$) of a trail formed by an Gaussian-shaped energy inclusion (peak value = $5.6 k_BT$, standard deviation = 4 pixels (px)) travelling at a speed of 0.05 px/$t_s$ towards the right. b) Average spin profile from the centre of the inclusion, corresponding to different time scales, as indicated by the color-coded vertical dashed lines in a). 
}
\end{figure}

To further confirm our claim that the trails are preferentially-adsorbed water-rich layers being advected downstream and explore an alternative mechanism, we simulated the 2D conserved order parameter (COP) Ising model \cite{newman1999monte}. For a direct analogue to a bacterium swimming in a critical fluid, we simulated a Gaussian-shaped high-energy region traversing the lattice at $\epsilon \approx 0.23$ and constant speed \cite{Demery2010}. In the co-moving frame, we see a downstream trail, Fig.~\ref{fig:Fig8}(a), whose transverse profile $n(y)$ shows a peak that decays oscillatorily, Fig.~\ref{fig:Fig8}(b), because of order parameter conservation, which is the direct analogue of mass conservation in our experiments. Experimentally, the situation is more complex: the observed oscillations are partly artefacts of phase-contrast imaging \cite{Elliot2001}.

The high shear rate near rotating flagella, $\dot\gamma \sim 10^4\,\si{\per\second}$ \cite{martinez2014flagellated}, might also generate a phase-bright trail in our system. If $\dot\gamma^{-1} \lesssim \tau$, the decay time of the fluctuations, we may expect the flow to homogenise the fluctuations \cite{Onuki1997}, reducing $\xi$ and therefore scattered light. The advection of a homogenised region downstream generates a phase-bright trail. Indeed, shear melting by flagella is responsible for bacterial trails in liquid crystals \cite{Zhou2014}.                                                                                                                                                                                                                     
 
\begin{figure}[t]
\includegraphics[width=0.5\textwidth,clip]{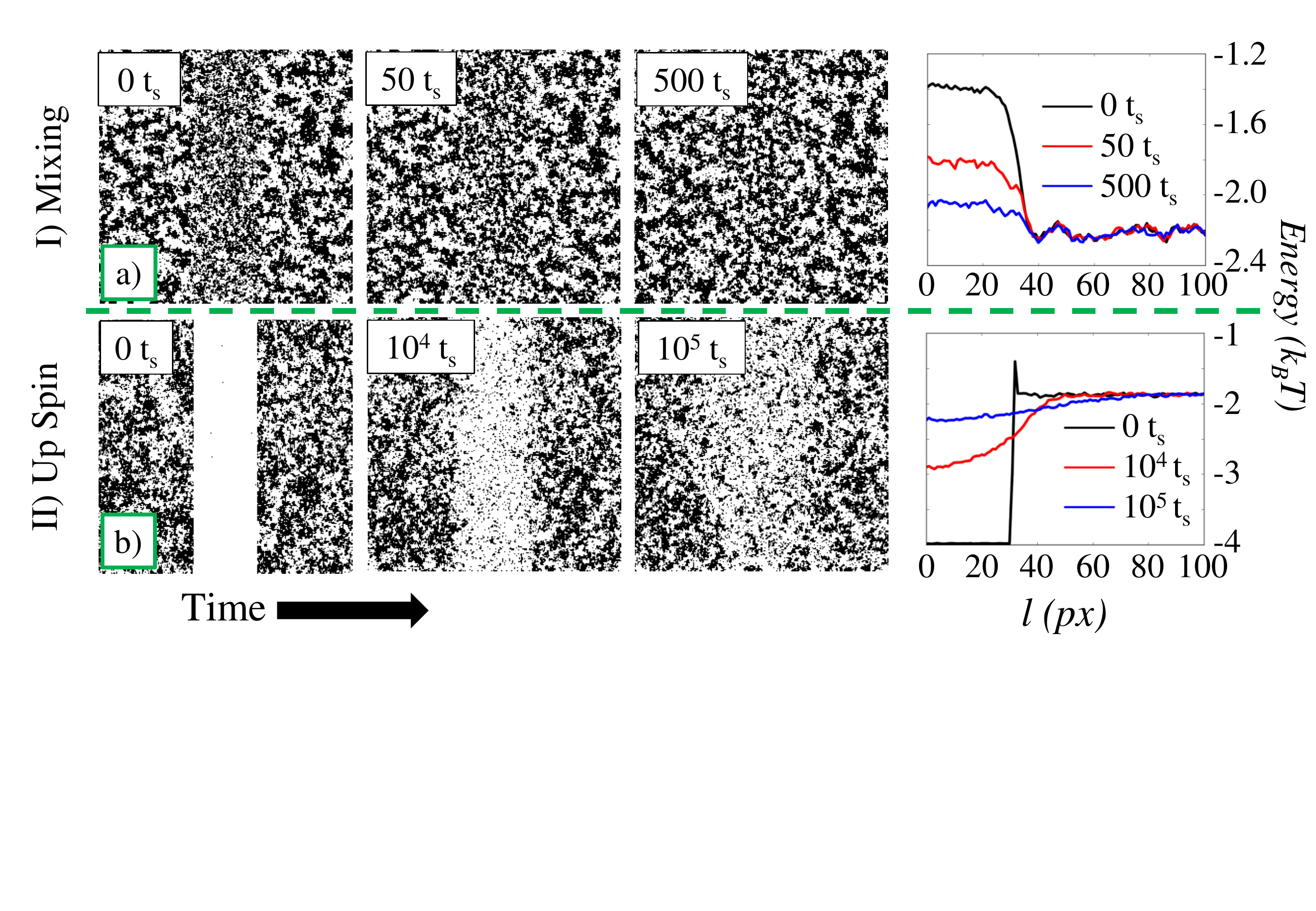}
\vspace{-2cm}
\caption{\label{fig:Fig9} 
Simulations of 64-site wide `trails' with different initial conditions embedded in a COP Ising lattice at $T_{\rm Bulk}=2.7$ ($\epsilon \approx 0.19$). Snapshots are shown at time $t = 0$ and two other times in units of Monte Carlo time step, $t_s$, and plots of the average energy as a function of the distance from the centre of the profile ($l$). a) A stripe initially at a higher temperature, $T = 4$ ($\epsilon \approx 0.76$). b) A stripe initially with all up spins at $T =2.7$. 
}
\end{figure}

To explore this possibility, we simulated the evolution of two kinds of stripes embedded in a near-critical COP Ising lattice ($\epsilon \approx 0.035, n= 0$). To mimic a shear-homogenised region, we used a stripe initially at a much higher temperature ($\epsilon \approx 0.76$), evidenced by the finer-scale fluctuations in the stripe compared to the bulk, Fig.~\ref{fig:Fig9}(a). Visually, this region becomes barely distinguishable from the bulk within $\tau_{\rm R}^{\rm shear} \sim 500t_s$, where $t_s$ is the MC time step. Plots of the energy profile of the stripe, $H(l)$, at increasing times verify this observation.

To mimic a water-rich trail, we prepared a stripe of positive spins at $\epsilon \approx 0.19$. Visual inspection and the $H(l)$ plot show that relaxation now takes $\sim 10^5 t_s \gtrsim 10^2 \tau_{\rm R}^{\rm shear}$. This is because a homogenised trail has the bulk concentration ($n = 0$), but a stripe of up spins has a different concentration ($n = 1$). Relaxation of the latter requires long-range transport subject to conserved (Kawasaki) dynamics. Thus, much slower differential adsorption effects will mask any shear effects in our system. 

The COP Ising model belongs to the same universality class as binary fluids with identical static critical exponents. Nevertheless, we do not attempt a more quantitative comparison between simulations and experiments, since dynamical critical phenomena and exponents are model dependent \cite{Hohenberg1977,Moser2006} and sensitive to details, e.g. the presence or absence of hydrodynamics. Thus, we would not expect the diffusivity exponents to be the same. {\color{black} However, the main conclusion we draw from these simulations, viz., the large separation in time scales between the relaxation processes in Fig.~\ref{fig:Fig9}(a) mimicking shear-homogenisation and in Fig.~\ref{fig:Fig9}(b) mimicking differential adsorption, should be valid, because this is traceable back to a fundamental, model-independent, distinction between the dynamics of processes requiring local reorganisation and long-range diffusive transport.}


To summarise, we have found that \textit{E. coli} bacteria swimming in a critical mixture of water and the surfactant \ce{C5E12} leave transient trails. That these trails are bright in phase contrast imaging together with analytic modelling and simulations suggest strongly that they are due to a preferentially-adsorbed water-rich layer on cell surfaces of mesoscopic extent due to proximity to criticality. This is advected downstream as cells swims. The same basic physics -- preferential affinity near criticality, underlies critical Casimir forces between colloids in critical fluid mixtures \cite{hertlein2008direct,GambassiPhysRevE,NellenEPL,vasilyev2014critical}, and gives trails in simulations of passive particles being dragged through a critical fluid \cite{Furukawa2003,Demery2010}. The underlying diffusive phenomenology behind the trail dissipation strongly suggests that the trails are not due to shear from flagella.

Finally, active matter dispersed in complex passive media can act as a probe of the latter. Thus, e.g., fast spinning bacterial flagella can be seen as a probe of the local high-shear rheology of polymer solutions on the nano-scale \cite{martinez2014flagellated}. Here, we have shown that self-propelled bacteria can be used to measure the dynamical critical exponent for diffusivity via fitting of the transverse intensity profile of our trails. Future experiments would no doubt demonstrate other examples of such use of active matter as local probes. 

\section{Acknowledgements} 

We thank J. Arlt, A. Brown, B. Guy, V. Martinez and T. Vissers for discussions. We additionally acknowledge two anonymous referees whose comments allowed for a significant improvement of this work. NK was part-funded by the EU (H2020-MSCA-IF-2014, ActiDoC No. 654688). All received funding from UK EPSRC (EP/J007404/1) and ERC (Advanced Grant ERC-2013-AdG 340877-PHYSAP).



\section{References} 

\begin{thebibliography}{10}
\providecommand*{\bibinfo}[2]{#2}
\providecommand*{\eprint}[1]{#1}
\providecommand*{\url}[1]{#1}
\bibitem{Ramaswamy2010}
\bibinfo{author}{S.~Ramaswamy}, \bibinfo{journal}{Ann. Rev. Condens. Matter
  Phys.} \bibinfo{volume}{\textbf{1}}(1), \bibinfo{pages}{323}
  (\bibinfo{date}{2010}).
\bibitem{FermiColloids}
\bibinfo{author}{W.~C.~K. Poon}, in \bibinfo{editors}{C.~Bechinger,
  F.~Sciortino, and P.~Ziherl}, eds., \emph{Physics of Complex Colloids}
  (\bibinfo{publisher}{IOS Press}, Amsterdam, \bibinfo{year}{2013}),
  \bibinfo{pages}{pp. 317--386}.
\bibitem{CatesReview2012}
\bibinfo{author}{M.~E. Cates}, \bibinfo{journal}{Rep. Prog. Phys.}
  \bibinfo{volume}{\textbf{75}}, \bibinfo{pages}{042601}
  (\bibinfo{date}{2012}).
\bibitem{Cates2016}
\bibinfo{author}{J.~Stenhammar}, \bibinfo{author}{R.~Wittkowski},
  \bibinfo{author}{D.~Marenduzzo}, and \bibinfo{author}{M.~E. Cates},
  \bibinfo{journal}{Sci. Adv.} \bibinfo{volume}{\textbf{2}},
  \bibinfo{pages}{e1501850} (\bibinfo{date}{2016}).
\bibitem{Arlt2018}
\bibinfo{author}{J.~Arlt}, \bibinfo{author}{A.~M. Martinez},
  \bibinfo{author}{A.~Dawson}, \bibinfo{author}{T.~Pilizota}, and
  \bibinfo{author}{W.~C.~K. Poon}, \bibinfo{journal}{Nature Comm.}
  \bibinfo{volume}{\textbf{9}}, \bibinfo{pages}{768} (\bibinfo{date}{2018}).
\bibitem{martinez2014flagellated}
\bibinfo{author}{V.~A. Martinez}, \bibinfo{author}{J.~Schwarz-Linek},
  \bibinfo{author}{M.~Reufer}, \bibinfo{author}{L.~G. Wilson},
  \bibinfo{author}{A.~N. Morozov}, and \bibinfo{author}{W.~C.~K. Poon},
  \bibinfo{journal}{Proc. Natl. Acad. Sci. (USA)}
  \bibinfo{volume}{\textbf{111}}, \bibinfo{pages}{17771}
  (\bibinfo{date}{2014}).
\bibitem{brown2016swimming}
\bibinfo{author}{A.~T. Brown}, \bibinfo{author}{I.~D. Vladescu},
  \bibinfo{author}{A.~Dawson}, \bibinfo{author}{T.~Vissers},
  \bibinfo{author}{J.~Schwarz-Linek}, \bibinfo{author}{J.~S. Lintuvuori}, and
  \bibinfo{author}{W.~C.~K. Poon}, \bibinfo{journal}{Soft Matter}
  \bibinfo{volume}{\textbf{12}}, \bibinfo{pages}{131} (\bibinfo{date}{2016}).
\bibitem{Vladescu2014}
\bibinfo{author}{I.~D. Vladescu}, \bibinfo{author}{E.~J. Marsden},
  \bibinfo{author}{J.~Schwarz-Linek}, \bibinfo{author}{V.~A. Martinez},
  \bibinfo{author}{J.~Arlt}, \bibinfo{author}{A.~N. Morozov},
  \bibinfo{author}{D.~Marenduzzo}, \bibinfo{author}{M.~E. Cates}, and
  \bibinfo{author}{W.~C.~K. Poon}, \bibinfo{journal}{Phys. Rev. Lett.}
  \bibinfo{volume}{\textbf{113}}, \bibinfo{pages}{268101}
  (\bibinfo{date}{2014}).
\bibitem{Lushi2014}
\bibinfo{author}{E.~Lushi}, \bibinfo{author}{H.~Wioland}, and
  \bibinfo{author}{R.~E. Goldstein}, \bibinfo{journal}{Proc. Natl. Acad. Sci.
  USA} \bibinfo{volume}{\textbf{111}}, \bibinfo{pages}{9733}
  (\bibinfo{date}{2014}).
\bibitem{Zhou2014}
\bibinfo{author}{S.~Zhou}, \bibinfo{author}{A.~Sokolov}, \bibinfo{author}{O.~D.
  Lavrentovich}, and \bibinfo{author}{I.~S. Aranson}, \bibinfo{journal}{Proc.
  Natl. Acad. Sci. (USA)} \bibinfo{volume}{\textbf{111}}, \bibinfo{pages}{1265}
  (\bibinfo{date}{2014}).
\bibitem{Abbott2014}
\bibinfo{author}{P.~C. Mushenheim}, \bibinfo{author}{R.~R. Trivedi},
  \bibinfo{author}{H.~H. Tuson}, \bibinfo{author}{D.~B. Weibel}, and
  \bibinfo{author}{N.~L. Abbott}, \bibinfo{journal}{Soft Matter}
  \bibinfo{volume}{\textbf{10}}, \bibinfo{pages}{88} (\bibinfo{date}{2014}).
\bibitem{Croze}
\bibinfo{author}{O.~A. Croze}, \bibinfo{author}{G.~P. Ferguson},
  \bibinfo{author}{M.~E. Cates}, and \bibinfo{author}{W.~C.~K. Poon},
  \bibinfo{journal}{Biophys. J.} \bibinfo{volume}{\textbf{101}},
  \bibinfo{pages}{525} (\bibinfo{date}{2011}).
\bibitem{Galajda}
\bibinfo{author}{P.~Galajda}, \bibinfo{author}{J.~Keymer},
  \bibinfo{author}{P.~Chaikin}, and \bibinfo{author}{R.~Austin},
  \bibinfo{journal}{J. Bact.} \bibinfo{volume}{\textbf{189}},
  \bibinfo{pages}{8704} (\bibinfo{date}{2007}).
\bibitem{Wioland2016}
\bibinfo{author}{H.~Wioland}, \bibinfo{author}{F.~G. Woodhouse},
  \bibinfo{author}{J.~Dunkel}, and \bibinfo{author}{R.~E. Goldstein},
  \bibinfo{journal}{Nature Phys.} \bibinfo{volume}{\textbf{12}},
  \bibinfo{pages}{341} (\bibinfo{date}{2016}).
\bibitem{buttinoni2012active}
\bibinfo{author}{I.~Buttinoni}, \bibinfo{author}{G.~Volpe},
  \bibinfo{author}{F.~K{\"u}mmel}, \bibinfo{author}{G.~Volpe}, and
  \bibinfo{author}{C.~Bechinger}, \bibinfo{journal}{J. Phys.: Condens. Matter}
  \bibinfo{volume}{\textbf{24}}, \bibinfo{pages}{284129}
  (\bibinfo{date}{2012}).
\bibitem{Buttinoni2013}
\bibinfo{author}{I.~Buttinoni}, \bibinfo{author}{J.~Bialk\'e},
  \bibinfo{author}{F.~K\"ummel}, \bibinfo{author}{H.~L\"owen},
  \bibinfo{author}{C.~Bechinger}, and \bibinfo{author}{T.~Speck},
  \bibinfo{journal}{Phys. Rev. Lett.} \bibinfo{volume}{\textbf{110}},
  \bibinfo{pages}{238301} (\bibinfo{date}{Jun 2013}).
\bibitem{Roy2016}
\bibinfo{author}{S.~Roy}, \bibinfo{author}{S.~Dietrich}, and
  \bibinfo{author}{F.~H{\"o}fling}, \bibinfo{journal}{J. Chem. Phys.}
  \bibinfo{volume}{\textbf{145}}, \bibinfo{pages}{134505}
  (\bibinfo{date}{2016}).
\bibitem{schwarz2016escherichia}
\bibinfo{author}{J.~Schwarz-Linek}, \bibinfo{author}{J.~Arlt},
  \bibinfo{author}{A.~Jepson}, \bibinfo{author}{A.~Dawson},
  \bibinfo{author}{T.~Vissers}, \bibinfo{author}{D.~Miroli},
  \bibinfo{author}{T.~Pilizota}, \bibinfo{author}{V.~A. Martinez}, and
  \bibinfo{author}{W.~C.~K. Poon}, \bibinfo{journal}{Colloids Surf. B}
  \bibinfo{volume}{\textbf{137}}, \bibinfo{pages}{2} (\bibinfo{date}{2016}).
\bibitem{hamano1985dynamics}
\bibinfo{author}{K.~Hamano}, \bibinfo{author}{T.~Sato},
  \bibinfo{author}{T.~Koyama}, and \bibinfo{author}{N.~Kuwahara},
  \bibinfo{journal}{Phys. Rev. Lett.} \bibinfo{volume}{\textbf{55}},
  \bibinfo{pages}{1472} (\bibinfo{date}{1985}).
\bibitem{hamano1995static}
\bibinfo{author}{K.~Hamano}, \bibinfo{author}{K.~Fukuhara},
  \bibinfo{author}{N.~Kuwahara}, \bibinfo{author}{E.~Ducros},
  \bibinfo{author}{M.~Benseddik}, \bibinfo{author}{J.~Rouch}, and
  \bibinfo{author}{P.~Tartaglia}, \bibinfo{journal}{Phys. Rev. E}
  \bibinfo{volume}{\textbf{52}}, \bibinfo{pages}{746} (\bibinfo{date}{1995}).
\bibitem{sinn1992}
\bibinfo{author}{C.~Sinn} and \bibinfo{author}{D.~Woermann},
  \bibinfo{journal}{Berichte der Bunsengesellschaft für physikalische Chemie}
  \bibinfo{volume}{\textbf{96}}, \bibinfo{pages}{913} (\bibinfo{date}{1992}).
\bibitem{Pollock}
\bibinfo{author}{L.~Emmerling}, \bibinfo{title}{\emph{Pollock}}
  (\bibinfo{publisher}{Taschen}, \bibinfo{year}{2016}).
\bibitem{Furukawa2003}
\bibinfo{author}{A.~Furukawa}, \bibinfo{author}{A.~Gambassi},
  \bibinfo{author}{S.~Dietrich}, and \bibinfo{author}{H.~Tanaka},
  \bibinfo{journal}{Phys. Rev. Lett.} \bibinfo{volume}{\textbf{111}},
  \bibinfo{pages}{055701} (\bibinfo{date}{2013}).
\bibitem{Demery2010}
\bibinfo{author}{V.~D\'emery} and \bibinfo{author}{D.~S. Dean},
  \bibinfo{journal}{Phys. Rev. Lett.} \bibinfo{volume}{\textbf{104}},
  \bibinfo{pages}{080601} (\bibinfo{date}{2010}).
\bibitem{Martinez2017Clem}
\bibinfo{author}{I.~A. Martinez}, \bibinfo{author}{C.~Devailly},
  \bibinfo{author}{A.~Petrosyan}, and \bibinfo{author}{S.~Ciliberto},
  \bibinfo{journal}{Entropy} \bibinfo{volume}{\textbf{19}}(77)
  (\bibinfo{date}{2017}).
\bibitem{Giavazzi2016}
\bibinfo{author}{F.~Giavazzi}, \bibinfo{author}{A.~Fornasieri},
  \bibinfo{author}{A.~Vailati}, and \bibinfo{author}{R.~Cerbino},
  \bibinfo{journal}{Eur. Phys. J. E} \bibinfo{volume}{\textbf{39}}(10),
  \bibinfo{pages}{103} (\bibinfo{date}{Oct 2016}).
\bibitem{newman1999monte}
\bibinfo{author}{M.~E.~J. Newman} and \bibinfo{author}{G.~T. Barkema},
  \bibinfo{title}{\emph{Monte Carlo Methods in Statistical Physics}}
  (\bibinfo{publisher}{Oxford University Press: New York, USA},
  \bibinfo{year}{1999}).
\bibitem{Kawasaki1966}
\bibinfo{author}{K.~Kawasaki}, \bibinfo{journal}{Phys. Rev.}
  \bibinfo{volume}{\textbf{145}}, \bibinfo{pages}{224} (\bibinfo{date}{May
  1966}).
\bibitem{Chen2000}
\bibinfo{author}{B.-H. Chen}, \bibinfo{author}{C.~A. Miller},
  \bibinfo{author}{J.~M. Walsh}, \bibinfo{author}{P.~B. Warren},
  \bibinfo{author}{J.~N. Ruddock}, \bibinfo{author}{P.~R. Garrett},
  \bibinfo{author}{F.~Argoul}, and \bibinfo{author}{C.~Leger},
  \bibinfo{journal}{Langmuir} \bibinfo{volume}{\textbf{16}},
  \bibinfo{pages}{5276} (\bibinfo{date}{2000}).
\bibitem{hertlein2008direct}
\bibinfo{author}{C.~Hertlein}, \bibinfo{author}{L.~Helden},
  \bibinfo{author}{A.~Gambassi}, \bibinfo{author}{S.~Dietrich}, and
  \bibinfo{author}{C.~Bechinger}, \bibinfo{journal}{Nature}
  \bibinfo{volume}{\textbf{451}}, \bibinfo{pages}{172} (\bibinfo{date}{2008}).
\bibitem{GambassiPhysRevE}
\bibinfo{author}{A.~Gambassi}, \bibinfo{author}{A.~Macio\l{}ek},
  \bibinfo{author}{C.~Hertlein}, \bibinfo{author}{U.~Nellen},
  \bibinfo{author}{L.~Helden}, \bibinfo{author}{C.~Bechinger}, and
  \bibinfo{author}{S.~Dietrich}, \bibinfo{journal}{Phys. Rev. E}
  \bibinfo{volume}{\textbf{80}}, \bibinfo{pages}{061143} (\bibinfo{date}{Dec
  2009}).
\bibitem{NellenEPL}
\bibinfo{author}{U.~Nellen}, \bibinfo{author}{L.~Helden}, and
  \bibinfo{author}{C.~Bechinger}, \bibinfo{journal}{EPL}
  \bibinfo{volume}{\textbf{88}}, \bibinfo{pages}{26001} (\bibinfo{date}{2009}).
\bibitem{vasilyev2014critical}
\bibinfo{author}{O.~A. Vasilyev}, \bibinfo{journal}{Physical Review E}
  \bibinfo{volume}{\textbf{90}}, \bibinfo{pages}{012138}
  (\bibinfo{date}{2014}).
\bibitem{ecology}
\bibinfo{author}{A.~Okubo} and \bibinfo{author}{S.~A. Levin},
  \bibinfo{title}{\emph{Diffusion and Ecological Problems: Modern
  Perspectives}} (\bibinfo{publisher}{Springer}, New York,
  \bibinfo{year}{2001}).
\bibitem{DrescherPuller}
\bibinfo{author}{K.~Drescher}, \bibinfo{author}{R.~E.Goldstein},
  \bibinfo{author}{N.~Michel}, \bibinfo{author}{M.~Polin}, and
  \bibinfo{author}{I.~Tuval}, \bibinfo{journal}{Phys. Rev. Lett.}
  \bibinfo{volume}{\textbf{105}}, \bibinfo{pages}{168101}
  (\bibinfo{date}{2010}).
\bibitem{giavazzi2016equilibrium}
\bibinfo{author}{F.~Giavazzi}, \bibinfo{author}{A.~Fornasieri},
  \bibinfo{author}{A.~Vailati}, and \bibinfo{author}{R.~Cerbino},
  \bibinfo{journal}{Eur. Phys. J. E} \bibinfo{volume}{\textbf{39}},
  \bibinfo{pages}{103} (\bibinfo{date}{2016}).
\bibitem{Elliot2001}
\bibinfo{author}{M.~S. Elliot} and \bibinfo{author}{W.~C. Poon},
  \bibinfo{journal}{Adv. Colloid Interface Sci.} \bibinfo{volume}{\textbf{92}},
  \bibinfo{pages}{133 } (\bibinfo{date}{2001}).
\bibitem{Onuki1997}
\bibinfo{author}{A.~Onuki}, \bibinfo{journal}{J. Phys.: Condens. Matter}
  \bibinfo{volume}{\textbf{9}}(29), \bibinfo{pages}{6119}
  (\bibinfo{date}{1997}).
\bibitem{Hohenberg1977}
\bibinfo{author}{P.~C. Hohenberg} and \bibinfo{author}{B.~I. Halperin},
  \bibinfo{journal}{Rev. Mod. Phys.} \bibinfo{volume}{\textbf{49}},
  \bibinfo{pages}{435} (\bibinfo{date}{1977}).
\bibitem{Moser2006}
\bibinfo{author}{R.~Folk} and \bibinfo{author}{G.~Moser}, \bibinfo{journal}{J.
  Phys. A: Math. Gen.} \bibinfo{volume}{\textbf{39}}, \bibinfo{pages}{R207}
  (\bibinfo{date}{2006}).

\end{thebibliography}
\end{document}